\documentclass{svjour3}
\smartqed
\usepackage{diagbox}
\usepackage{amssymb}
\usepackage{amsmath}
\usepackage{graphicx}
\usepackage{subfigure}
\usepackage{setspace}
\usepackage{multirow}

\journalname{Quantum Information Processing}
\begin{document}
\title{Analyses and improvement of a broadcasting multiple blind signature scheme based on quantum GHZ entanglement}

\author{Wei Zhang \and Daowen Qiu \and Xiangfu Zou \and Paulo Mateus}

\institute{Wei Zhang \and Daowen Qiu \at
              Institute of Computer Science Theory, School of Data and Computer Science,  Sun Yat-sen University, Guangzhou 510006, China \\
              The Guangdong Key Laboratory of Information Security Technology, Sun Yat-sen
University, Guangzhou 510006, China\\
              \email{issqdw@mail.sysu.edu.cn (Corresponding author's address)}\\
\and
Wei Zhang  \at
              School of Mathematics and Statistics, Qiannan Normal College for Nationalities, Duyun 558000, China \\
              \and
 Xiangfu Zou  \at
              School of Mathematics and Computational Science, Wuyi University, Jiangmen 529020, China \\
               \and Paulo Mateus \at
 SQIG--Instituto de Telecomunica\c{c}\~{o}es, Departamento de Matem\'{a}tica, Instituto Superior T\'{e}cnico, Av. Rovisco Pais
              1049-001, Lisbon, Portugal}
\maketitle
\begin{abstract}
A broadcasting multiple blind signature scheme based on quantum GHZ entanglement has been presented recently. It is said that the scheme's unconditional security is guaranteed by adopting quantum key preparation, quantum encryption algorithm and quantum entanglement. In this paper, we prove that each signatory can get the signed message just by an intercept-resend attack. Then, we show there still exists some participant attacks and external attacks. Specifically, we verify the message sender Alice can impersonate each signatory to sign the message at will, and so is the signature collector Charlie. Also, we demonstrate that the receiver Bob can forge the signature successfully, and with respect to the external attacks, the eavesdropper Eve can modify the signature at random. Besides, we discover Eve can change the signed message at random, and  Eve can impersonate Alice as the message sender without being discovered. In particular, we propose an improved scheme based on the original one and show that it is secure against not only the attacks mentioned above but also some collusion attacks.

\end{abstract}

\section{Introduction}

Quantum signature is the counterpart in the quantum world of classical digital signature. Compared with the classical one, quantum digital signature is based on the laws of quantum physics, which makes it own many natural advantages in the aspect of security. Therefore, quantum digital signature has foreseeable application in E-payment system, E-business and E-government.

In 2001, Gottesman and Chuang \cite{2} proposed a quantum digital signature scheme based on a quantum one-way function and quantum swap test. After that,
much progress has been made. Zeng and Keitel \cite{3} presented an arbitrated quantum signature scheme by using GHZ entanglement in 2002. In 2009, Li et al \cite{4} designed a more efficient arbitrated quantum signature scheme by using Bell state. Zou  and Qiu \cite{5} proposed an arbitrated quantum signature without entanglement in 2010. Along with the development of quantum signature, more and more quantum signature models have been proposed for different application demands, such as quantum proxy signature \cite{6,7,8,9,10}, quantum group signature \cite{11,12,13,14,15}, quantum blind signature \cite{16,17,18,19,20} and quantum multiple signature \cite{21,22,23}.

A secure quantum signature scheme should satisfy two basic requirements:(1) No forgery. Exactly speaking, the signature cannot be forged by any illegal signatory.(2) No disavowal. The signatory cannot disavow his signature and the receiver cannot disavow his receiving the signature and its integrity\cite{4}.

Gao et al.  \cite{24} presented a perfect cryptanalysis on existing arbitrated quantum signature. They pointed out that the signature can be forged by the receiver in almost all the existing arbitrated quantum signature (AQS) schemes. Zou and Qiu gave some attacks and corresponding improvements of fair quantum blind signature schemes \cite{25}. After that, Lin et al. further pointed out that there still exists a secure leakage caused by the reuse of signing key in the fair quantum blind signature schemes \cite{26}.  In view of the existence of these serious loopholes, it is imperative to reexamine the security of other quantum signature protocols.

Recently, a broadcasting multiple signature scheme based on quantum GHZ entanglement has been proposed in Ref. \cite{1}. It could be used to settle the problem that a message is so important that needs to be signed by multiple signatories, in order to guarantee the message's privacy, none of signatories can acquire what they have signed. Maybe it can be applied in E-bank system. For example: A large number of money has to be transferred through E-bank system on the internet. The E-bank system operator submits the request to the bank after filling the application form including payment amount, bank transfer account and some other information. When the request arrives, the bank clerk signs to approve. But it is not enough, it has to ask the manager's authority, then it needs to be signed by the manager. In the whole process, all the signatories cannot learn what they have signed. But the application form has been recorded in the E-bank system, when disagreement takes place, the bank can track the message sender.

In the original work, it is said that the scheme's unconditional security is guaranteed by adopting quantum key preparation, quantum encryption algorithm and quantum entanglement.  Here we show that each signatory can get the message that is to be signed just by an intercept-resend attack. Furthermore, we verify there still exists some participant attacks and external attacks. Specifically, we discover the message sender Alice can impersonate $U_{i}$ to sign the message, and so is the signature collector Charlie. Additionally, we demonstrate the receiver Bob can forge the signature successfully, and with respect to the external attacks, the eavesdropper Eve can modify the signature at random. Besides, we find Eve can change the message that is to be signed at will, and Eve can impersonate Alice as the message sender without being discovered. Finally, we particularly design an improved scheme based on the original one, and show that the new scheme can resist the attacks that the original scheme are encountered mentioned above, and it can also resist some collusion attacks.

The rest of this paper is organized as follows. First, in Section 2, we briefly review the original scheme. In Section 3, we present the attack strategies of the original scheme in detail. Particularly, in Section 4 we design an improved scheme based on the original one. Then in Section 5, we make a security analysis of the improved scheme. Finally, in section 6 we  make a short conclusion and give some future issues.

\section{Original scheme}

\subsection{Preliminary}
A qubit $|\psi\rangle$ is expressed as a vector in two-dimensional Hilbert Space. Generally, $\{|0\rangle, |1\rangle\}$ is a group of typical orthonormal basis, which is called  $Z$-basis. However, there still exists another group of orthonormal basis called $X$-basis, denoted as $\{|+\rangle, |-\rangle\}$, where

\begin{eqnarray}
&& |+\rangle = \frac{|0\rangle+|1\rangle}{\sqrt{2}}
\end{eqnarray}
and
\begin{eqnarray}
&& |-\rangle = \frac{|0\rangle-|1\rangle}{\sqrt{2}}.
\end{eqnarray}
From Eq. (1) and (2), it is easy to get
\begin{eqnarray}
&& |0\rangle = \frac{|+\rangle+|-\rangle}{\sqrt{2}}
\end{eqnarray}
and
\begin{eqnarray}
&& |1\rangle = \frac{|+\rangle-|-\rangle}{\sqrt{2}}.
\end{eqnarray}
Then, a single particle state $|\psi\rangle$ can be written in $Z$-basis as
\begin{eqnarray}
&& |\psi\rangle = a|0\rangle+b|1\rangle
\end{eqnarray}
satisfying
\begin{eqnarray}
&& |a|^{2}+|b|^{2}=1.
\end{eqnarray}
According to Eqs. (3) and (4), it can also be expressed as
\begin{eqnarray}
&& |\psi\rangle = \frac{a+b}{\sqrt{2}}|+\rangle+\frac{a-b}{\sqrt{2}}|-\rangle.
\end{eqnarray}

The original scheme is mainly based on GHZ entanglement state, which is a three-particle maximum entanglement state expressed as
\begin{eqnarray}
&& |\phi\rangle = \frac{|0_{A}0_{B}0_{C}\rangle+|1_{A}1_{B}1_{C}\rangle}{\sqrt{2}}.
\end{eqnarray}
Meanwhile, it can also be expressed in $X$-basis as
\begin{eqnarray}
&& |\phi\rangle = \frac{1}{2}(|+,+,+\rangle_{ABC}+|+,-,-\rangle_{ABC}+|-,+,-\rangle_{ABC}+|-,-,+\rangle_{ABC}).
\end{eqnarray}
 By Eq. (9), it is showed that the state of the particle $C$ can be deduced by measuring the particles $A$ and $B$ in the $X$-basis respectively. In other words, the state of any particle can be deduced  if  the other two particles are determined. For example, if  particle $A$ and $B$ are in the state of $|+\rangle$, then particle $C$ will be $|+\rangle$ definitely. We show the correlation of GHZ state in Table 1.

\begin{table}[h]
\center
\begin{tabular}{|l|c|c|}
 \hline
 \diagbox{\quad \quad  A  \quad}{ \quad \quad C  \quad}{ \quad \quad B  \quad} & \quad \quad \quad $|+\rangle_{B}$ \quad \quad \quad& \quad \quad \quad$|-\rangle_{B}$\quad \quad \quad  \\
 \hline
 \quad \quad \quad$|+\rangle_{A}$\quad \quad \quad
& \quad \quad \quad $|+\rangle_{C}$ \quad \quad \quad & \quad \quad \quad $|-\rangle_{C}$ \quad \quad \quad
\\
\hline
 \quad \quad \quad$|-\rangle_{A}$\quad \quad \quad & \quad \quad \quad$|-\rangle_{C}$\quad \quad \quad
& \quad \quad \quad$|+\rangle_{C}$\quad \quad \quad  \\
 \hline
 \end{tabular}
 \caption{\quad \quad \quad \quad \quad \quad \quad \quad \quad \quad \quad Correlation of GHZ state.}
 \center
 \end{table}

\subsection{The scheme}

 the original scheme involves four characters: (1) Alice is the message sender. (2) $U_{i}$ $(i=1, 2, \cdots, t)$ is $i$-th member of broadcasting multiple signatory. (3) Charlie is the signature collector. (4) Bob is the receiver and verifier.

The scheme is composed of four parts: initial phase, the individual blind signature generation and verification phase, the combined multiple signature phase and the combined multiple blind signature verification phase.

In original scheme, Alice sends $t$ copies of an $n$-bit classical string $m$ to $t$ signatories $U_{i}$ $(i=1,2,\cdots,t)$, respectively. Then $U_{i}$ signs message $m$ to get the blind signature $S_{i}$ and sends it to Charlie. Charlie collects and verifies these blind signatures, then he constructs a multiple signature and sends it to Bob. Finally, Bob verifies the multiple signature by confirming the message.
\begin{enumerate}
\item Initial Phase

\begin{enumerate}

\item Alice transforms the  message $m$ into $n$-bit as $m = m(1)\|m(2)\| \cdots\| m(j)\|$ $ \cdots\| m(n)$. The message $m$ is to be signed bit by bit.

\item Quantum key distribution. Alice shares secret key $K_{AB}$ with Bob, secret keys $K_{AU_{i}}$ $(i=1,2,\cdots,t)$ with each signatory $U_{i}$ respectively, secret key $K_{AC}$ with Charlie. Charlie shares secret keys $K_{CU_{i}}$$(i=1,2,\cdots,t)$ with each signatory $U_{i}$ respectively. Bob shares secret key $K_{BC}$ with Charlie. To obtain unconditional security, all these keys are distributed via QKD protocols \cite{27,28}.

\item Alice sends $K_{AB}(m)$ to Bob. Here Alice encrypts $m$ into $K_{AB}(m)$ by using her own secret key $K_{AB}$ according to the one-time pad encryption algorithm. Specifically, $K_{AB}(m)=m\oplus K_{AB}$.
\end{enumerate}

\item The Individual Blind Signature Generation and Verification Phase

Here we just pick one of the signatory $U_{i}$ as the representative to make an illustration.

2.1  Quantum Channel Setup

Alice generates $n$ GHZ entanglement states which are in state of $|\phi\rangle_{ACU_{i}}$ denoted as $\{|\phi(1)\rangle_{ACU_{i}}, |\phi(2)\rangle_{ACU_{i}}, \cdots, |\phi(j)\rangle_{ACU_{i}}, \cdots, |\phi(n)\rangle_{ACU_{i}}\}$. Then Alice distributes the particle $C$ and $U_{i}$ of each GHZ state to Charlie and the signatory $U_{i}$ respectively.

2.2  Blind Signature and Its Verification

\begin{enumerate}

\item Alice measures her GHZ particle sequence in $X$-basis to get a classical string $a=\{a(1), a(2), \cdots, a(j), \cdots, a(n)\}$ according to
\begin{equation}
 a(j)  =
 \begin{cases}
    0 &  \text{if  \quad the measurement outcome is $+$}, \\
    1 &  \text{if  \quad the measurement outcome is $-$}.
 \end{cases}
 \end{equation}
Then Alice publishes the classical string $m^{*}$ as
\begin{eqnarray}
&& m^{*}=a\oplus m.
\end{eqnarray}
Note: Here we do some modifications based on the original work as the measurement cannot be performed according to the message $m$, but it still maintains the original work.

\item Alice encrypts $a$ by using the secret key $K_{AC}$ according to the one-time pad encryption algorithm and sends $K_{AC}(a)$ to Charlie.

\item Charlie measures his GHZ particles in the $X$-basis and records the measurement outcome sequence $c=\{c(1), c(2), \cdots, c(j), \cdots, c(n)\}$, where
\begin{equation}
 c(j)  =
 \begin{cases}
    0 &  \text{if  \quad the measurement outcome is $+$}, \\
    1 &  \text{if  \quad the measurement outcome is $-$}.
 \end{cases}
 \end{equation}

\item In order to provide the audit voucher, Charlie has to convert the measuring result $c$ by quantum fingerprinting function as follows:
\begin{eqnarray}
&& |f(x)\rangle = \frac{1}{\sqrt{m}}\sum_{i=1}^{m}|i\rangle|E_{i}(x)\rangle.
\end{eqnarray}
Then Charlie encrypts the result $|f(c)\rangle$ with the key $K_{CU_{i}}$  according to QOTP algorithm, resulting in
\begin{eqnarray}
&& |H\rangle = E_{K_{CU_{i}}}(|f(c)\rangle).
\end{eqnarray}
Here $E_{K_{CU_{i}}}$ is the quantum encryption algorithm for qubits \cite{29}. After that, Charlie sends $|H\rangle$ to $U_{i}$.

\item On receiving $|H\rangle$, $U_{i}$ measures his own GHZ particles to get the result $S_{i}$ according to
\begin{equation}
S_{i}(j)  =
 \begin{cases}
    0 &  \text{if  \quad the measurement outcome is $+$}, \\
    1 &  \text{if  \quad the measurement outcome is $-$}.
 \end{cases}
 \end{equation}
Then $U_{i}$ sends the encrypted result $K_{CU_{i}}(S_{i})$ to Charlie. Here $U_{i}$ encrypts $S_{i}$ into $K_{CU_{i}}(S_{i})$ according to one-time pad algorithm.

\item Charlie decrypts $K_{CU_{i}}(S_{i})$  into $S_{i}$ by using the secret key $K_{CU_{i}}$. Due to the string $c$ and the correlation of the GHZ state, Charlie can figure out Alice's measurement outcomes $a'$. Then Charlie can get the message $m'$ as
\begin{eqnarray}
&& m'=m^{*}\oplus a'.
\end{eqnarray}
Note that $a'$ and $m'$ will be equal to $a$ and $m$ respectively if there is no mistake happened in the communication process.

\item Charlie decrypts $K_{AC}(a)$ into $a$ by using his secret key $K_{AC}$, generates $m$ with $m^{*}$ and compares $m$ with $m'$. If they are equal, Charlie accepts $S_{i}$, otherwise, it is rejected.

\end{enumerate}

\item The Combined Multiple Signature Generation Phase

Charlie collects all individual signatures $S_{i}$ $(i=1,2,\cdots,i,\cdots,t)$ and generates the message $m'_{1}, m'_{2}, \cdots, m'_{i}, \cdots, m'_{t}$. If $m'_{i}=m'_{i+1}(i=1,2,\cdots,i,\cdots,t-1)$, he confirms the message and generates the multiple signature $S=S_{1}\|S_{2}\|\cdots\|$ $S_{i}\|\cdots\|S_{t}$, otherwise, he terminates the process. After confirming the message, Charlie  sends $K_{BC}(m'_{1})$ to Bob. Here $m'_{1}$ is turned into $K_{BC}(m'_{1})$ according to one-time pad algorithm.

\item The Multiple signature Verification Phase

Bob decrypts $K_{BC}(m'_{1})$ and $K_{AB}(m)$, and he accepts the signature if $m'_{1}=m$, otherwise, he terminates the process.
\end{enumerate}

\section{Attacks on Tian Yu's scheme}

In this section, we will show there are some participant attacks and external attacks in the scheme. Here we just take a signatory $U_{i}$ as a representative to illustrate the attack strategy in detail. Sometimes, we just take one bit of the message that is to be signed to make a demonstration.

\subsection{The signatory $U_{i}$ can get the message $m$}

In order to make a clear illustration of $U_{i}$'s attack strategy, we rewrite the GHZ entanglement state as follows:

\begin{eqnarray}
|\phi\rangle_{ACU_{i}} &&= \frac{|0_{A}0_{C}0_{U_{i}}\rangle+|1_{A}1_{C}1_{U_{i}}\rangle}{\sqrt{2}}\\\nonumber
&&= \frac{|0_{A}\rangle|0_{C}0_{U_{i}}\rangle+|1_{A}\rangle|1_{C}1_{U_{i}}\rangle}{\sqrt{2}} \\\nonumber
&&=\frac{|+\rangle_{A}}{\sqrt{2}}(\frac{|00\rangle_{_{CU_{i}}}+|11\rangle_{CU_{i}}}{\sqrt{2}})+\frac{|-\rangle_{A}}{\sqrt{2}}(\frac{|00\rangle_{_{CU_{i}}}-|11\rangle_{CU_{i}}}{\sqrt{2}}).
\end{eqnarray}

Next, we describe the signatory $U_{i}$'s intercept-resend attack strategy in detail. Firstly, $U_{i}$ intercepts the GHZ particle $C$ when it is sent from Alice to Charlie and combine it with his own GHZ particle $U_{i}$, then he performs a two particle measurement in Bell-basis. Then $U_{i}$ can deduce the state of GHZ particle $A$ according to the measurement outcomes. If the measurement outcome is $\beta_{00}$, according to Eq. (17),  particle $A$ is in the state of $|+\rangle_{A}$ definitely, then $U_{i}$ can further get $a(j)=0$. If the measurement outcome is $\beta_{10}$,  particle $A$ is in the state of $|-\rangle_{A}$  and get $a(j)=1$. Here
\begin{eqnarray}
&& |\beta_{00}\rangle_{CU_{i}}=\frac{|00\rangle_{CU_{i}}+|11\rangle_{CU_{i}}}{\sqrt{2}},\\
&& |\beta_{01}\rangle_{CU_{i}}=\frac{|01\rangle_{CU_{i}}+|10\rangle_{CU_{i}}}{\sqrt{2}},\\
&& |\beta_{10}\rangle_{CU_{i}}=\frac{|00\rangle_{CU_{i}}-|11\rangle_{CU_{i}}}{\sqrt{2}}
\end {eqnarray}
and
\begin{eqnarray}
 &&|\beta_{11}\rangle_{CU_{i}}=\frac{|01\rangle_{CU_{i}}-|10\rangle_{CU_{i}}}{\sqrt{2}}.
\end {eqnarray}
According to Eq. (11), $U_{i}$ can obtain $m(j)$ with the $m^{*}$ published by Alice in Step $2.2(a)$ . After that, $U_{i}$ resends the GHZ particle $C$ to Charlie. All of these cannot be discovered in the verifying phase.

\subsection{The signatory $U_{i}$ can get Charlie's measurement outcome $c$}

In original scheme,  Charlie's measurement outcome $c$ is encrypted by the quantum fingerprinting function according to Eq. (13) before sending it to $U_{i}$. Consequently, $U_{i}$ cannot get $c$ by decrypting it directly. In part 3.1, we have showed $U_{i}$ can get Alice's measurement result by intercept-resend attack, then $U_{i}$ can get $c$ based on the correlation of the GHZ state after he measures his GHZ particles $U_{i}$ in X-basis. Therefore, the encryption of $c$ is failed. Furthermore, state $|H\rangle$ sent from Charlie to $U_{i}$ in Step 2.2(d) is useless, then it can be removed.

\subsection{The message sender Alice can impersonate $U_{i}$ to sign message at will}

Here we show Alice can impersonate $U_{i}$ to sign message in the original scheme. In the signature phase, Alice sets up the quantum channel by generating $n$ GHZ entanglement states and then sending particle $C$ and $U_{i}$ to Charlie and signatory $U_{i}$ separately. In this step, Alice can send particle $U_{i}$ to the signatory but postpone to send particle $C$ to Charlie. Meanwhile, she measures the two particles in her hand in Bell-basis and records the measurement outcomes. According to Eq. (17), she can deduce the state of particle $U_{i}$ based on the measurement outcome, according to Eq. (15), she can get $U_{i}$'s signature $S_{i}$ . After that, Alice sends particle $C$ to Charlie.

In addition, Alice can get $U_{i}$'s secret key $K_{CU_{i}}$ by intercept-resend attack. Firstly, Alice intercepts $K_{CU_{i}}(S_{i})$ in Step 2.2(e). Then she can get $K_{CU_{i}}$ by adding $S_{i}$ to $K_{CU_{i}}(S_{i})$ as
\begin{eqnarray}
&& K_{CU_{i}}= S_{i} \oplus K_{CU_{i}}(S_{i}).
\end{eqnarray}
After that, Alice resends $K_{CU_{i}}(S_{i})$ to Charlie.

From above, we can see Alice can not only get $S_{i}$ but also the secret key $K_{CU_{i}}$, then Alice can impersonate $U_{i}$ successfully. Worse still, Alice can sign arbitrary message at will. Alice can intercept $K_{CU_{i}}(S_{i})$ and resend an arbitrary $K_{CU_{i}}(S'_{i})$ to Charlie, meanwhile, she modifies her measurement outcomes $a$ in Step 2.2(a) to satisfy the correlation of the GHZ state. Therefore, Alice's cheating behaviour cannot be discovered in the verification phase.

\subsection{The collector Charlie can impersonate $U_{i}$ successfully}

According to the original scheme, collector Charlie can get Alice's measurement outcome $a$ and his own outcome $c$, then he can deduce the state of particle $U_{i}$ based on the correlation of GHZ state. Therefore, he can get $U_{i}$'s signature $S_{i}$. Besides, Charlie has the secret key $K_{CU_{i}}$,  consequently, Charlie can impersonate $U_{i}$ successfully. Even more, Charlie can also sign the message at random. Exactly,  Charlie can discard $U_{i}$'s signature $S_{i}$, instead, he generates an arbitrary $S'_{i}$ and modifies his measurement outcome $c$ according to Table 1 to maintain the GHZ correlation. Then $S'_{i}$ can pass the verification process definitely.

\subsection{The receiver Bob can forge $U_{i}$'s signature}

In the original scheme, the signatory $U_{i}$ generates the blind signature $S_{i}$ by measuring his particle in X-basis according to Eq. (15). Here we show the receiver Bob can forge the signature by intercept-resend attack.

Firstly, the receiver Bob intercepts $K_{CU_{i}}(S_{i})$ when it is sent from $U_{i}$ to Charlie and add an $n$-bit random string
\begin{eqnarray}
&& l=i_{1}i_{2}\cdots i_{n}
\end{eqnarray}
to $K_{CU_{i}}(S_{i})$, then Charlie will get
\begin{eqnarray}
&& S'_{i}=S_{i}\oplus l.
\end{eqnarray}
In order to make sure $S'_{i}$ can pass the verification process, Bob also intercepts $K_{AC}(a)$, adds another $n$-bit random string
\begin{eqnarray}
&& l'=j_{1}j_{2}\cdots j_{n}
\end{eqnarray}
to $K_{AC}(a)$ and resends it to Charlie. Then Charlie will get
\begin{eqnarray}
&& a''=a\oplus l'
\end{eqnarray}
instead of $a$.

Next, we illustrate that Bob can figure out $l'$ based on $l$ and the correlation of GHZ state as follows:

\begin{enumerate}
\item  If $S_{i}(j)=0$, then we can infer that the state of particle $U_{i}$ is $|+\rangle$. From Table 1, we can see both of particle $A$ and $C$ are in state of $|+\rangle$ or in state of $|-\rangle$. In other words, $a(j)=c(j)=0$ or $a(j)=c(j)=1$.

\item  If $S_{i}(j)=1$, then particle $U_{i}$ is the state of $|-\rangle$. According to Table 1, particle $A$ and $C$ are in the state of $|+\rangle$ and $|-\rangle$ or $|-\rangle$ and $|+\rangle$ respectively. That is to say $a(j)=0, c(j)=1$ or $a(j)=1, c(j)=0$.

\end{enumerate}
From above, we can find that
\begin{eqnarray}
&& S_{i}(j)\oplus a(j)\oplus c(j)=0
\end{eqnarray}
is satisfied in both of the two cases. Therefore, if $S'_{i}$ can pass the verification, according to Eq. (27),  $S'_{i}(j)$, $a'(j)$ and $c(j)$ are bound to satisfy
\begin{eqnarray}
&& S'_{i}(j)\oplus a''(j)\oplus c(j)=0.
\end{eqnarray}
Then we can get
\begin{eqnarray}
&& l(j)\oplus l'(j)=0.
\end{eqnarray}
Therefore, we can easily get $l=l'$.

After that, Bob adds $l$ to the message $m$ which is received from Alice in Step 1(c) in the initial phase, according to the scheme, $S'_{i}$ will be accepted as $U_{i}$'s blind signature of message $m\oplus l$. Therefore, Bob can forge the signature successfully.

\subsection{The eavesdropper Eve can change the message $m$ at will}

Firstly, we show Eve can get message $m$ by intercept-resend attack. Eve can intercept GHZ particle $U_{i}$ and $C$ when they are sent from Alice to $U_{i}$ and Charlie separately. Then she measures them in Bell-basis, according to Eq. (17), Eve can get each $a(j)$ based on her own measurement outcome. According to Eq. (11), Eve can get $m$ with $m^{*}$ published by Alice.

Next, we show Eve can get Alice's secret key $K_{AC}$ and $K_{AB}$. Eve also can get Alice's secret key $K_{AB}$ by intercept-resend method. Eve  intercepts $K_{AB}(m)$ when it is sent from Alice to Bob, then she can get $K_{AB}$ by adding the message $m$ to $K_{AB}(m)$ as
\begin{eqnarray}
&& K_{AB}=K_{AB}(m)\oplus m.
\end{eqnarray}
Meanwhile, Eve can compute $a$ by using $m^{*}$ published by Alice in Step 2.2(a). Similarly, Eve can get $K_{AC}$ using the same method.

From above, we can see Eve can not only get the message $m$ but also Alice's secret keys, then Eve can impersonate Alice as the message sender. Besides, Eve can intercept $K_{AC}(a)$ and $K_{AB}(m)$ and resend another pair of $K_{AC}(a')$ and $K_{AB}(m'')$ to Charlie and Bob respectively, satisfying                                                                                                                                                                                 $m^{*}=a'\oplus m''$. According to the oringinal scheme, message $m$ will be changed into $m''$ and this modification cannot be discovered in the verification process. As $m''$ is arbitrary, then Eve can change the  message $m$ at will.

\subsection{The eavesdropper Eve can modify the signature at will}

Eve can intercept the GHZ particle $U_{i}$ and $C$ when they are sent from Alice to $U_{i}$ and Charlie separately. Instead, she performs a Pauli operator $Z$ on each particle and then sends them to $U_{i}$ and Charlie separately. Next we show Eve can change the signature through this method.

Assume Alice's measurement outcome is $a(j)=0$, according to Eq. (10), GHZ particle $A$ is in the state of $|+\rangle$. From Table 1, we can see particle $U_{i}$ and $C$ are in two different cases: Case 1: both of them are in state of $|+\rangle$ and Case 2: both of them are in state of $|-\rangle$.
Next, we show that no matter what case it is, the signature will be modified under Eve's attack and this modification can pass the verification process.
\begin{enumerate}

\item Case 1:
\begin{enumerate}

\item Without Eve's attack. In this occasion, we can easily see that $U_{i}$ will generate $S_{i}(j)=0$ and Charlie will get $c(j)=0$ by measuring their own particle in X-basis respectively.

\item Under Eve's attack. The state of particle $U_{i}$ is changed from $|+\rangle$ to $Z|+\rangle=|-\rangle$, so is particle $C$. We can get  $S'_{i}(j)=1$ and $c'_{i}(j)=1$, but $S'_{i}(j)$, $c'(j)$ and $a(j)$ still satisfy
\begin{eqnarray}
&& S'_{i}(j) \oplus c'(j) \oplus a(j)=0.
\end{eqnarray}
Then $S'_{i}(j)$ can pass the verification process.

\end{enumerate}

\item Case 2 can be presented similarly.
\end{enumerate}
From the above, we can see Eve can modify the signature at will.
\section{An improved scheme}

In this section, we design an improved scheme based on the original one. Before presenting the new scheme, it is necessary to introduce the QOTP algorithm utilized in this paper. Suppose a quantum message
\begin{align}
|P\rangle=\bigotimes^{l}_{j=1}|P_{j}\rangle
\end{align}
is composed of $l$ qubits
\begin{align}
|P_{j}\rangle=\alpha_{j}|0\rangle+\beta_{j}|1\rangle,
\end{align}
where
\begin{align}
|\alpha_{j}|^{2}+|\beta_{j}|^{2}=1.
\end{align}
 The QOTP encryption $E_{K}$ used in this scheme can be described as
\begin{eqnarray}
&& E_{K}(|P\rangle)=\bigotimes^{l}_{j=1} \sigma^{K_{4j}}_{x}\sigma^{K_{4j-1}}_{z}T\sigma^{K_{4j-2}}_{x}\sigma^{K_{4j-3}}_{z}|P_{j}\rangle
\end{eqnarray}
where
\begin{align}
W =\frac{i}{\sqrt{3}}(\sigma_{x}-\sigma_{y}+\sigma_{z}).
\end{align}
This QOTP encryption algorithm is firstly introduced in Ref. [30]. The assistant operator $W$ can promise the encrypted message not to be forged. Specifically, for arbitrary quantum message $|P\rangle$, there are no non-identity unitary operator $V$ and $U$ such that
\begin{align}
E^{\dag}_{K}VE_{K}|P\rangle \equiv U |P\rangle.
\end{align}
Assuming that there are a couple of non-identity unitary operators $U$ and $V$ satisfying Eq. (37), then message $|P\rangle$ can be modified into $U |P\rangle$ deterministically by the attacker in its transmission even though $|P\rangle$ has been encrypted into $E_{K}|P\rangle$ according to QOTP algorithm. Specifically, when $|P\rangle$ has been encrypted into $E_{K}|P\rangle$ and transmitted in the quantum channel, the attacker Eve can intercept $E_{K}|P\rangle$ and perform the unitary operator $V$ on it and resend $V E_{K}|P\rangle$ to the receiver, thus the receiver performs the decryption operator $E^{\dag}_{K}$ on $V E_{K}|P\rangle$ after he receives it.  According to  Eq. (37), the receiver will finally get $U|P\rangle$ in stead of $|P\rangle$. For more details we can refer to \cite{24,30}. Introducing the improved QOTP algorithm into our new protocol is mainly to avoid this problem.

In part 3.4 we can see the collector Charlie can alter the individual signature $S_{i}$ at random in the original scheme. In order to make sure the originality of signature generated by each signatory $U_{i}$ in the improved scheme, we define  a one-way hash function  \cite{31}:
 \begin{eqnarray}
&&H(x): \{0, 1\}^{*}\longrightarrow \{0, 1\}^{n}.
\end{eqnarray}

After introducing the improved QOTP and defining the hash function, we pay attention to the GHZ entanglement. Firstly,  we rewrite GHZ state $|\phi\rangle$ as
 \begin{eqnarray}
|\phi\rangle=\frac{|000\rangle_{123}+|111\rangle_{123}}{\sqrt{2}}
 =\frac{|+\rangle_{1}\otimes|\beta_{00}\rangle_{23}}{\sqrt{2}}+\frac{|-\rangle_{1}\otimes|\beta_{10}\rangle_{23}}{\sqrt{2}}.
\end{eqnarray}
From Eq. (39), we can see if the particle $1$ is in the state $|+\rangle$, then the particles $2$ and $3$ will be in the state $|\beta_{00}\rangle$ definitely. Similarly, if the particle $1$ is observed to be $|-\rangle$, then the particles $2$ and $3$ will be $|\beta_{10}\rangle$. Next, we do three operations on the GHZ state $|\phi\rangle$ as follows:
\begin{enumerate}
\item Perform a measurement on the particle $1$ in X-basis and record the measurement outcomes according to
\begin{equation}    a_{1} =
 \begin{cases}
    0  &  \text{if the outcome is  $+$}, \\
    1  &  \text{if the outcome is  $-$}.
 \end{cases}
 \end{equation}
\item Perform a Pauli operator $I$ or $X$ randomly on the particle $2$ and record the operation as
\begin{equation}    b_{1} =
 \begin{cases}
    0  &  \text{if the operator is  $I$}, \\
    1  &  \text{if the operator is  $X$}.
 \end{cases}
 \end{equation}
\item Do a two particle measurement on the particles $2$ and $3$ in Bell basis and record the outcomes as
\begin{equation}    c_{1} =
\begin{cases}
    00  &  \text{if the state is observed as $|\beta_{00}\rangle$}, \\
    01  &  \text{if the state is observed as $|\beta_{01}\rangle$}, \\
    10  &  \text{if the state is observed as $|\beta_{10}\rangle$}, \\
    11  &  \text{if the state is observed as $|\beta_{11}\rangle$}.
\end{cases}
\end{equation}
\end{enumerate}
Then we can find that
 \begin{eqnarray}
&& c_{1}=a_{1}\|b_{1}
\end{eqnarray}
is always satisfied. This will be utilized in our new scheme later.

Our new scheme involves $t+3$ participants, namely the message sender Alice, $t$ signatories $U_{1},U_{2},\cdots, U_{t}$, the signature collector Charlie and the verifier Bob. Firstly, Alice prepares $t$ copies of $n$-bit classical message $m$ and conceals each of them with corresponding secret keys shared before, and then she sends the blind messages to each signatory $U_{i}$. Subsequently, each $U_{i}$ signs the blind message to generate individual signature and sends it to Charlie. On receiving all the individual signatures, Charlie verifies each individual signature and aggregates them into a multi-signature. Finally, Bob verifies the validity of the multi-signature.
\\
\\
\par

\begin{picture}(150,150)
\put(70,145){(1)}
\put(190,145){(2)}
\put(280,145){(3)}

\setlength{\unitlength}{1.5cm}
\put(0.5,1.3){\framebox(0.6,0.5){Alice}}

\put(2.5,0.1){\framebox(0.6,0.4){U$_{t}$}}
\put(2.5,0.8){\framebox(0.6,0.4){$\vdots$}}
\put(2.5,1.8){\framebox(0.6,0.4){U$_{2}$}}
\put(2.5,2.6){\framebox(0.6,0.4){U$_{1}$}}

\put(4.5,1.3){\framebox(0.8,0.5){Charlie}}

\put(6.5,1.3){\framebox(0.6,0.5){Bob}}

\put(0.25,-0.2){\dashbox{0.2}(3.1,3.4){}}
\put(3.6,-0.2){\dashbox{0.2}(2.0,3.4){}}
\put(5.9,-0.2){\dashbox{0.2}(1.6,3.4){}}

\put(1.45,2.9){1}
\put(1.4,1.77){\vector(0,3){1.02}}
\put(1.2,1.88){\vector(0,3){0.92}}
\put(1.6,1.22){\vector(0,3){1.57}}
\put(1.8,0.78){\vector(0,3){2.01}}
\put(2.72,1.55){3}
\put(2.72,1.28){5}
\put(3.42,0.05){7}
\put(5.7,1.55){4}
\put(3.8,0.81){\line(0,3){1.99}}
\put(4.0,1.3){\line(0,3){1.5}}
\put(4.28,1.7){\line(0,3){1.09}}
\put(4.40,1.81){\line(0,3){0.99}}
\put(3.8,2.8){\vector(1,0){1.3}}
\put(5.3,2.7){2}
\put(3.9,0.5){\line(0,3){1.74}}
\put(4.1,0.5){\line(0,3){1.30}}
\put(4.28,0.5){\line(0,3){0.82}}
\put(4.40,0.5){\line(0,3){0.72}}
\put(3.9,0.5){\vector(1,0){1.2}}
\put(5.3,0.4){6}
\put(1.18,2.8){\line(1,0){0.64}}

\put(1.1,1.8){\vector(4,3){1.4}}
\put(1.1,1.3){\vector(4,-3){1.4}}
\put(1.1,1.7){\vector(4,1){1.4}}
\put(1.1,1.35){\vector(4,-1){1.4}}
\put(1.1,1.5){\vector(1,0){3.4}}
\put(4.5,1.47){\vector(-1,0){3.4}}

\put(3.1,2.8){\vector(4,-3){1.4}}
\put(3.1,2.0){\vector(4,-1){1.4}}
\put(3.1,1.0){\vector(3,1){1.4}}
\put(3.1,0.3){\vector(4,3){1.4}}
\put(4.5,1.8){\vector(-4,3){1.4}}
\put(4.5,1.7){\vector(-4,1){1.4}}
\put(4.5,1.3){\vector(-4,-3){1.4}}
\put(4.5,1.4){\vector(-3,-1){1.4}}

\put(5.3,1.5){\vector(1,0){1.2}}

\put(0.8,1.3){\vector(0,-1){1.3}}
\put(0.8,0.0){\vector(1,0){6.0}}
\put(6.8,0.0){\vector(0,1){1.3}}

\end{picture}

\vskip 2mm
\vskip 2mm
\vskip 2mm
\vskip 2mm
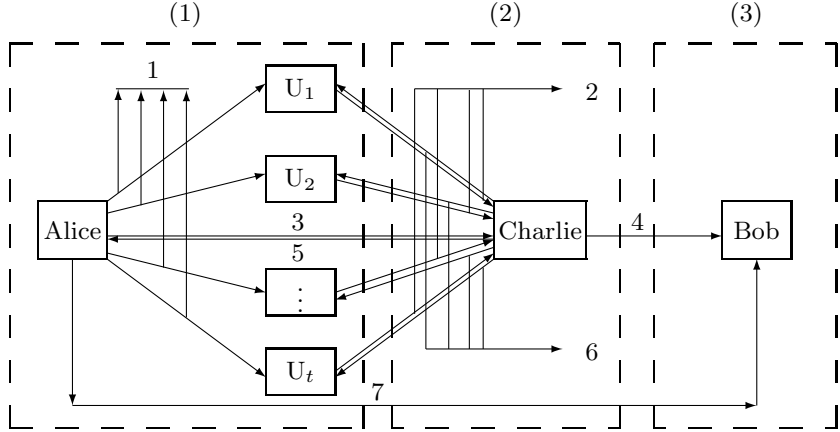
\centerline{\textbf{Figure 1.} The improved scheme: \text{ (1) individual blind signature phase;}}
\text{ (2) individual signature verification and multi-signature generation phase;}

\text{ (3) multi-signature verification phase; 1 $E_{K_{AU_{i}}}(|\psi(M_{i})\rangle)$; 2 $E_{K_{CU_{i}}}(|\psi(S_{i})\rangle)$ }

\text{  and $E_{K_{CU_{i}}}(|\psi(M'_{i})\rangle)$; 3 $E_{K_{AC}}(|\psi(a_{1})\rangle)$ and $E_{K_{AC}}(|\psi(T)\rangle)$; 4 $E_{K_{BC}}(|\psi(S)\rangle)$ }

\text{  and $E_{K_{BC}}(|\psi(m')\rangle)$; 5 $|\phi\rangle_{1}$; 6 $|\phi\rangle_{2}$; 7 $E_{AB}(|\psi(m)\rangle)$.}

\vskip 2mm
\par

The scheme is also composed of four phases: the initial phase, the individual blind signature generation phase, the individual signatures verification and the multi-signature generation phase, and the multi-signature verification phase. The brief procedure of our scheme has been illustrated in Fig.1, and the description in detail is presented as follows.
\subsection{Initial phase}
\begin{enumerate}
\item Alice transforms the original message into $n$-bit sequence as
\begin{eqnarray}
&& m=m(1)\| m(2)\|\cdots\| m(n).
\end{eqnarray}
Message $m$ is signed bit by bit.
\item Quantum key distribution. Alice shares $4n$-bit secret keys $K_{AB}$, $K_{AC}$ and $K_{AU_{i}}$ with Bob, Charlie and each signatory $U_{i}$, respectively. Charlie shares a $8n$-bit secret key $K_{CU_{i}}$  with each signatory $U_{i}$. Bob shares  a $4n$-bit secret key $K_{BC}$ with Charlie. In order to ensure unconditional security, all the keys are distributed by QKD protocols.
\item Alice transforms classical message $m$ into $n$-qubit state
\begin{eqnarray}
&& |\psi(m)\rangle=\bigotimes^{n}_{j=1}|\psi(m(j))\rangle
\end{eqnarray}
according to computational basis $\{|0\rangle, |1\rangle\}$ (i.e., $|\psi(m(j))\rangle = |0\rangle(|1\rangle)$, when $m(j)=0(1)$) and sends $E_{K_{AB}}(|\psi(m)\rangle)$ to Bob, where $E_{K_{AB}}$ is according to QOTP algorithm introduced above. Note that, in subsequent phase, all the classical information is turned into quantum states and encrypted by the same QOTP algorithm before transmission.
\end{enumerate}

\subsection{ The individual blind signature generation phase}
\begin{enumerate}
\item Message blinding and transmission. Alice prepares $t$ copies of $n$-bit classical message $m$ and blinds it into
\begin{eqnarray}
&& M_{i}=m\oplus K_{AB}^{(n)}\oplus K_{AU_{i}}^{(n)}
\end{eqnarray}
where $K_{AB}^{(n)}$ and $K_{AU_{i}}^{(n)}$ are the first $n$-bit of the secret keys $K_{AB}$ and $K_{AU_{i}}$ respectively. Then she sends $E_{K_{AU_{i}}}(|\psi(M_{i})\rangle)$ to each signatory $U_{i}$. After that she also generates
\begin{eqnarray}
&& T=m\oplus\bigoplus_{i=1}^{t}M_{i}
\end{eqnarray}
and sends $E_{K_{AC}}(|\psi(T)\rangle)$ to Charlie.
\item Quantum channel setup. Charlie prepares $n$ GHZ states $|\phi\rangle$ denoted as
\begin{eqnarray}
&& |\phi\rangle=\bigotimes^{n}_{j=1}|\phi(j)\rangle,\\
&& |\phi(j)\rangle=\frac{|000\rangle_{123}+|111\rangle_{123}}{\sqrt{2}}
\end{eqnarray}
and sends the first and second particles of each GHZ state to Alice and each signatory $U_{i}$ respectively, keeping the third ones to his own. We use $|\phi\rangle_{1}$, $|\phi\rangle_{2}$ and $|\phi\rangle_{3}$ to denote the states of the first, second and third particles sequence:
\begin{eqnarray}
&& |\phi\rangle_{l}=\bigotimes^{n}_{j=1}|\phi(j)\rangle_{l}, l=1,2,3.
\end{eqnarray}
Note that all the particles are distributed via secure quantum channel here. Otherwise, we should add an entanglement checking process to make sure the entanglement is maintained during the whole signature process.
\item Alice's measurement. Alice generates an $n$-bit stochastic string $a_{1}$ by performing a measurement on $|\phi\rangle_{1}$ in X-basis according to
\begin{equation}    a_{1}(j) =
 \begin{cases}
    0  &  \text{if the state is observed as $|+\rangle$}, \\
    1  &  \text{if the state is observed as $|-\rangle$}.
 \end{cases}
 \end{equation}
Then she sends $E_{KC}(|\psi(a_{1})\rangle)$ to Charlie.
\item Individual signature generation. At this point, we use $U_{i}$ as a representative to make a demonstration. First of all, $U_{i}$ gets the blind message $M'_{i}$ by first decrypting and then measuring in computational basis when he receives $E_{K_{AU_{i}}}(|\psi(M_{i})\rangle)$ from Alice. Next, he generates its signature $S_{i}$. In our new scheme, each individual signature $S_{i}$ is a $2n$-bit random string which is composed of two parts: valid part and auxiliary part. The auxiliary part is used to ensure the valid part's originality during their transmission. We denote it as
\begin{eqnarray}
&& S_{i}=S_{i}^{(1)}\|S_{i}^{(2)},\\
&& S_{i}^{(2)}=H(R_{i}\|S_{i}^{(1)}\|M'_{i}),\\
&& R_{i}=K_{AU_{i}}\oplus K _{CU_{i}}^{(4n)}.
\end{eqnarray}
On receiving each $|\phi(j)\rangle_{2}$, each $U_{i}$ generates the valid part $S_{i}^{(1)}$  by performing a unitary operator $I$ or $X$ on each $|\phi(j)\rangle_{2}$ randomly:
\begin{equation}    S_{i}(j) =
 \begin{cases}
    0  &  \text{if $U_{i}$ chooses to perform $I$}, \\
    1  &  \text{if $U_{i}$ chooses to perform $X$}.
 \end{cases}
 \end{equation}
Then $U_{i}$ sends $E_{K_{CU_{i}}^{(4n)}}(|\phi'\rangle_{2})$ to Charlie.
\end{enumerate}

\subsection{ The individual blind signatures verification and the multi-signature generation phase}

\begin{enumerate}
\item Charlie gets the string $a_{1}'$ and $T'$. First of all, Charlie gets $a_{1}'$ and $T'$ by performing a measurement on $|\psi(a_{1})\rangle$ and $|\psi(T)\rangle$ in computational basis respectively after decrypting $E_{AC}(|\psi(a_{1})\rangle)$ and $E_{AC}(|\psi(T)\rangle)$ on receiving them from Alice.
\item Charlie generates a $2n$-bit random string $c_{1}$. Charlie combines each $|\phi'(j)\rangle_{2}$ with his own particle $|\phi(j)\rangle_{3}$ to form a two particle state after decrypting $E_{K_{CU_{i}}^{(4n)}}(|\phi'\rangle_{2})$. Then he performs a two particle measurement in Bell basis to generate a $2n$-bit random string $c_{1}$ according to
\begin{equation}    c_{1}(2j-1)c_{1}(2j) =
\begin{cases}
    00  &  \text{if the state is observed as $|\beta_{00}\rangle$}, \\
    01  &  \text{if the state is observed as $|\beta_{01}\rangle$}, \\
    10  &  \text{if the state is observed as $|\beta_{10}\rangle$}, \\
    11  &  \text{if the state is observed as $|\beta_{11}\rangle$}.
\end{cases}
\end{equation}
\item Charlie gets $S'_{i}$ and $M''_{i}$. After getting $a_{1}'$ and $c_{1}$, Charlie asks $U_{i}$ to send $E_{K_{CU_{i}}}(|\psi(S_{i})\rangle)$ and $E_{K_{CU_{i}}}(|\psi(M'_{i})\rangle)$  to him. Then he measures $|\psi(S_{i})\rangle$ and $|\psi(M'_{i})\rangle$ in computational basis to abstract $S'_{i}$ and $M''_{i}$ after decrypting them.

\item Verification process of the individual signature $S_{i}$. Owning to $a_{1}'$, $c_{1}$ and $S'_{i}$, Charlie verifies $S_{i}$ by verifying
\begin{eqnarray}
&& c_{1}(2j-1)c_{1}(2j)=a_{1}'(j)S_{i}^{'(1)}(j), \quad ( j=1,2,\cdots,n)
\end{eqnarray}
is satisfied or not. If it is satisfied, $S'_{i}$ is accepted by Charlie as $U_{i}$'s signature of blind message $M''_{i}$, then he stores the pair $(M''_{i}, S'_{i})$. Otherwise, $S'_{i}$ is rejected by Charlie.
\item Multi-signature generation. Assume that $S'_{1}, S'_{2}, \cdots, S'_{t}$ have been generated and verified by Charlie, then Charlie produces the multi-signature $S$ as
\begin{eqnarray}
&& S =  \bigoplus_{i=1}^{t}S_{i}^{'(1)}.
\end{eqnarray}
At the same time, Charlie creates $T''$ by
\begin{eqnarray}
&& T'' =  \bigoplus_{i=1}^{t}M''_{i}.
\end{eqnarray}
Then he can produce the message $m'$ through
\begin{eqnarray}
&& m' =  T''\oplus T'.
\end{eqnarray}
$S$ is generated by Charlie as the multi-signature of $m'$. After that, Charlie sends $E_{BC}(|\psi(m')\rangle)$ and $E_{BC}(|\psi(S)\rangle)$ to Bob.

\end{enumerate}

\subsection{The multi-signature verification phase}
\begin{enumerate}
\item Bob verifies the message $m$. Bob abstracts the  message $m'$ and $m''$ by performing a measurement on $|\psi(m)\rangle$ and $|\psi(m')\rangle$ in basis of $\{|0\rangle, |1\rangle\}$ respectively. Then he compares them with each other. If $m'=m''$, Bob publishes the verification parameter $V_{1}=1$ and continues to carry out the following steps. Otherwise, he publishes $V_{1}=0$ and terminates the scheme.

\item Bob verifies the multi-signature. After affirming the parameter $V_{1}=1$, Alice announces each $M_{i}$ $(i=1, 2, \cdots, t)$ and Charlie announces each $S'_{i}$ on the public board. Meanwhile, each signatory $U_{i}$ publishes the string $R_{i}$ which is used to generate their signature $S_{i}$. On receiving all the information, Bob abstracts the muti-signature $S'$ by performing a measurement on $|\psi(S)\rangle $ in computational basis. Then Bob verifies whether the following equations are satisfied or not:
\begin{eqnarray}
&& S'=\bigoplus_{i=1}^{t}S_{i}^{'(1)},\\
&& S_{i}^{'(2)}=H(R_{i}\|S_{i}^{'(1)}\|M_{i}),\quad (i=1, 2, \cdots, t).
\end{eqnarray}
If all the equations are satisfied, Bob accepts $S'$ as the multi-signature of $m'$. Otherwise, he rejects it and aborts the  scheme.
\end{enumerate}

Finally, we list our improvements as follows:
\begin{enumerate}
\item All the classical information is transformed into quantum message before transmission. Meanwhile, it is encrypted according to the improved QOTP algorithm which is introduced above.
\item  Each individual blind signature is generated by performing a random operation on a GHZ particle rather than measuring it directly.
\item  The GHZ entanglement can be maintained during the whole signature process by using secure quantum channel.
\item  The originality of each individual signature can be ensured by utilizing a hash function. Additionally, each blinded message $M'_{i}$ is used to generate a component of the individual signature $S'_{i}$ according to Eq. (53) which ensures that any disturbance of the blinded message will destroy the signature scheme.
\item Public board is utilized in the verification process which ensures that everyone can perform the verification when all the information is published.
\item The size of the multi-signature is constant rather than the original scheme which is linear with the number of signatory.
\end{enumerate}
Unfortunately, our new scheme's security is based on the utilized hash function rather than unconditional security.

\section{Security analysis}
In this section, we analyze the security of the new scheme. As we know, a secure signature scheme should satisfy no forgery and no disavowal. Because our scheme is a blind multiple signature which owns the merit of both blind signature and multiple signature at the same time, we should also talk about the blindness and the traceability. Blindness indicates the signatory cannot know the content of the message that he has signed \cite{32}. Traceability means once disagreement takes place, the signatory can trace the message owner \cite{32}. Additionally, we show that the new scheme is secure against some collusion attack. Collusion attack is a kind of attack strategy that some dishonest participants may collude to do some cheating such as forging the signature without other participants' participation or denying what they have done in the signing phase \cite{33,34,35}.

\subsection{No forgery}
\subsubsection{Alice cannot forge the signature}
Each individual signature $S_{i}$ is generated by the signatory $U_{i}$'s performing a Pauli operator $I$ or $X$ on his own GHZ particle sequence randomly. Therefore, Alice cannot get any information on each individual signature rather than guessing. As a result, Alice has to do some cheating in the signature's transmission to forge the signature successfully. Maybe there are two opportunities. One is that Alice performs the forgery attack when the individual signature $S_{i}$ is transmitted from $U_{i}$ to Charlie. Unfortunately, all the classical information is transformed into quantum states and encrypted according to the improved QOTP algorithm first proposed in Ref. \cite{30} in the new scheme. It is said that any quantum message encrypted by the QOTP algorithm cannot be forged. Then the forgery attack will get failed definitely. The other opportunity is to utilize the GHZ correlation existing among Alice, $U_{i}$ and Charlie. Through this method, Alice has to control the whole quantum entanglement channel. Unfortunately, this cannot be realized as the entanglement is distributed by secure quantum channel. Thus, this attack strategy is bound to fail. Briefly, Alice cannot forge an arbitrary individual signature. Similarly, it is impossible for Alice to forge the multi-signature.

\subsubsection{Charlie cannot forge the signature}

Charlie, the signature collector who can get all the individual signatures and generate the multi-signature, is considered to be most likely to forge the signature successfully. Here we show that Charlie cannot forge the signature either. Because Charlie can get each individual signature and generate the multi-signature, he can forge the signature by modifying each individual signature $S'_{i}$ into $S''_{i}$ and keeping the message $m'$ unaltered. As a result, the original multi-signature $S'$ is changed into $S''$. Charlie sends $S''$ instead of $S'$ to Bob as the signature of $m'$. Charlie's forgery attack seems to be successful, but Charlie's dishonest behavior is to be caught in the verification process because  Eq. (62) cannot be satisfied. Charlie can modify each $S_{i}^{'(1)}$ randomly, but he cannot know how to alter the corresponding $S_{i}^{'(2)}$ to fit his modification because $R_{i}$ is only owned by $U_{i}$ before it is published. From the above, we can see it is impossible for Charlie to forge the signature.

\subsubsection{Bob cannot forge the signature}
Bob, the receiver and verifier, can forge the signature by substituting another $S''$ for the actual $S'$ after it has been verified. Then he claims that $S''$ is the signature of the message $m'$. Here we show Bob's forgery attack will get failed because everyone can witness his dishonest behavior by verifying Eq.(61) with all the individual signatures being announced on the public board.

\subsubsection{No forgery under participants' collusion attack}
The single participant's forgery attacks have been discussed above, so we begin to  talk on participants' collusion attacks:
\begin{enumerate}

\item The collusion among partial signatories.

To make a clear illustration, we assume that the first $t-1$ signatories collaborate to forge the multi-signature $S$ in this paper. In order to forge the multi-signature $S$ successfully, they have to bypass $U_{t}$ and forge the individual signature $S_{t}$. According to the scheme, $S_{t}$ is a $2n$-bit string composed of $S_{t}^{(1)}$ and $S_{t}^{(2)}$. $S_{t}^{(1)}$ is generated by $U_{t}$'s performing a Pauli operator $I$ or $X$ on his GHZ particle sequence randomly and then it is transmitted after being turned into quantum message and then being encrypted by the improved QOTP algorithm. The other $t-1$ signatories cannot acquire it other than guessing. Even though they can guess $S_{t}^{(1)}$ correctly by a fluke, their forgery attack will get failed as they cannot get $R_{t}$ and $M_{t}$ to generate the corresponding $S_{t}^{(2)}$ to pass the verification. Consequently, partial signatories cannot forge the signature.

\item The collusion between partial signatories and Alice.

Partial signatoriesing with Alice can get $M_{t}$ but still cannot get $U_{t}$'s $R_{t}$, so they cannot forge the signature either.

\item The collusion between partial signatories and Charlie.

$S_{t}$ is sent from $U_{t}$ to Charlie, then they can get $U_{t}$'s individual signature. Here we mainly show they cannot modify $S_{t}$. Charlie can modify $S_{t}^{(1)}$, meanwhile, he modifies the corresponding string $c$ to satisfy Eq. (57), then this modification can pass the individual signature verification process. Unfortunately, the modification cannot pass the multi-signature verification process. Though Charlie has the blind message $M'$ and the modified $S_{t}^{''(1)}$, they are still lack of $U_{t}$'s personal string $R_{t}$ to alter $S_{t}^{'(2)}$ to fit the modified $S_{t}^{''(1)}$. Therefore, their dishonest behavior will be discovered definitely.

\item The collusion between partial signatories and Bob.

Partial signatories choose to collaborate with Bob, they can get the message $m'$ and derive $U_{t}$'s individual signature $S'_{t}$, but they cannot modify $S'_{t}$ because they do not have the essential material $M'_{t}$ and $R_{t}$.

\item The collusion between Alice and Charlie.

Charlie in cooperation with Alice can ensure him to get each blind message $M_{i}$ before published, but this cannot make them to forge the signature successfully because of the absence of $R_{i}$.

\end{enumerate}

\subsection{No disavowal}
\subsubsection{Each signatory cannot disavow his individual signature}
Each signatory cannot disavow the truth that they have signed the message because each individual signature $S_{i}$ contains the string $R_{i}$ including the secret keys $K_{AU_{i}}$ and $K_{CU_{i}}^{(4n)}$ which are only owned by $U_{i}$. After verification, $R_{i}$ has been published on the public board. If signatory $U_{i}$ disavows the signature for his own benefit, his dishonest will be caught by Alice and Charlie by verifying Eq. (54).

\subsubsection{Impossibility for Bob's disavowal}
Bob's disavowal includes that Bob disavows his receiving  or the integrity of the multi-signature. Firstly, we show Bob cannot disavow his receiving the signature. Bob should announce the verification parameter $V_{1}$ after checking the message, which indicates Bob has received $E_{BC}(|\psi(m')\rangle)$ from Charlie. According to the scheme, $E_{BC}(|\psi(S)\rangle)$ is sent with $E_{BC}(|\psi(m')\rangle)$ simultaneously, Bob cannot disavow his receiving the signature. Even if Bob sticks to that he has not got the signature, Charlie can send him $E_{BC}(|\psi(S)\rangle)$ again or even publishes $S$. Then everyone can witness he has received the signature. Next, we show Bob cannot disavow the signature's integrity. If $m'=m''$ but Bob claims that $m'\neq m''$  for his own benefit, we can ask Alice, Charlie and Bob to announce the message $m$ respectively. Then Bob's dishonest behavior will be discovered by Alice and Charlie according to the voting rule. Note that here we assume that Alice and Charlie are just loyal to their own and there is no collaborate attack.

\subsection{Secure against some external attacks}
In the previous section, we have showed that the eavesdropper Eve can forge the signature successfully by performing an intercept-resend attack on the original scheme. Here we show our new scheme is secure against some external attacks. First of all, we talk on the entanglement auxiliary particle attack. Entanglement auxiliary particle attack is a general strategy for entanglement based protocols. By this method, attackers entangle an ancillary particle into the entanglement state by a CNOT operation and then disentangle it from the obtained state by applying another CNOT operation to abstract what they want to know to forge the signature \cite{36}. Unfortunately, the GHZ entanglement particles are distributed through secure quantum channel in the new scheme, then the entanglement auxiliary particle cannot be attached. Therefore, this attack can be avoided. Next, we turn to the intercept-resend attack. All the classical information is transformed into quantum message and encrypted by the improved QOTP algorithm, so the intercept-resend attack will be failed. At last, we concern about the man-in-middle attack. Man-in-middle attack means the malicious attacker counterfeits the signatory and sends simultaneously particles and message to the receiver to temper the message or forge the signature \cite{32}. In the new scheme, secret keys distributed via QKD protocol are shared among all the participants. Owing to the unconditional security of QKD protocol, it is impossible for the malicious attacker to perform man-in-middle attack to temper the message and forge the signature.

\subsection{Blindness}
In the new scheme, the message sender Alice sends the blinded message $M_{i}=m\oplus K_{AB}^{(n)}\oplus K_{AU_{i}}^{(n)}$ to each signatory $U_{i}$ after being encrypted by the improved QOTP algorithm. As $U_{i}$ cannot get the secret key $K_{AB}$ shared between Alice and Bob, it is impossible for $U_{i}$ to abstract the message $m$.

\subsection{Traceability}
The new scheme is a kind of blind signature scheme, and therefore, each signatory $U_{i}$ cannot learn the content of the message. But $U_{i}$ can track the message owner when there is a disagreement taking place. As the blinded message $M_{i}=m\oplus K_{AB}^{(n)}\oplus K_{AU_{i}}^{(n)}$, it includes the components of the secret keys $K_{AB}$ and $K_{AU_{i}}$ simultaneously. This indicates the message is from Alice definitely because $K_{AB}$ and $K_{AU_{i}}$ are only owned simultaneously by Alice.

\section{Conclusion}

In this paper, we have analyzed the security of a broadcasting multiple blind signature scheme based on quantum GHZ entanglement. We have pointed out that there exists some participant attacks and external attacks in the scheme and the attack strategies have been presented in detail. After that, we have designed an improved scheme and showed that the new scheme is secure against the attacks that are encountered by the original scheme. Besides, the new scheme is secure against some collusion attack. Unfortunately, the security of our new scheme is based on the utilized hash function rather than unconditional security. Recently,  based on quantum homomorphic signature \cite{37},  an unconditional secure broadcasting blind multiple signature scheme has been designed. Maybe it has provided us some probability to design an unconditional secure one in the future. The secure quantum channel has been utilized in our new scheme, which will make it less practical. Fortunately, a practical quantum digital signature has been presented recently \cite{38}, in which the secure quantum channel has been removed. It is also worth considering to design a more practical scheme in the future. Additionally, an anonymous reviewer points out that the length of secret keys is much longer than the message, which makes the protocol less efficient. It is also worth to considering to improve it in the future.

\begin{acknowledgements}
The authors would like to thank the referees for their very helpful suggestions that greatly
helped to improve the quality of this paper. This work is supported in part by the National
Natural Science Foundation of China (Nos.  61572532, 61272058), the Natural Science Foundation of Qiannan Normal
College for Nationalities joint Guizhou Province of China (No. Qian-Ke-He LH Zi[2015]7719), the Natural Science Foundation of Central Government Special Fund for Universities of West China (No. 2014ZCSX17) and the
Foundation of Graduate Education Reform of Wuyi University (No. YJS-JGXM-14-02).

\end{acknowledgements}

\end{document}